\documentclass[a4paper,11pt]{article}
\pdfoutput=1 

\usepackage{jcappub} 

\usepackage[T1]{fontenc} 

\title{\boldmath Bayesian evidence of the post-Planck curvaton}

 \author{Robert J. Hardwick}
 \author{and Christian T. Byrnes}
\affiliation{Astronomy Centre, University of Sussex, Falmer, Brighton BN1 9QH, UK}

\emailAdd{r.hardwick@physics.org}
\emailAdd{c.byrnes@sussex.ac.uk}

\abstract{We perform a Bayesian model comparison for scenarios within the quadratic curvaton model, determining the degree to which both are disfavoured with respect to the $\Lambda$CDM concordance model and single-field quadratic inflation, using the recent \emph{Planck} data release.  Despite having three additional model parameters, the simplest curvaton scenario is not disfavoured relative to single-field quadratic inflation, and it becomes favoured against this single-field model when we include the joint BICEP/Keck/\emph{Planck} analysis.  In all cases we assume an instantaneous inflaton decay and no surviving isocurvature perturbations. Despite the success of \emph{Planck} reaching its forecast measurement accuracy, we show that the current constraints on local non-Gaussianity are insufficiently precise to have any significant impact on the evidence ratios so far. We also determine the precision $\sigma(f_{\mathrm{NL}})$ required by future measurements assuming a fiducial value of $f_{\mathrm{NL}}=-5/4$ or $10.8$ to no longer disfavour the curvaton against the  $\Lambda$CDM parametrisation, and we discuss the effect that the predicted increase in precision from future measurements on $f_{\mathrm{NL}}$ may have. We show that our results are not very sensitive to our choice of priors.}

\begin{document}
\maketitle
\flushbottom

\section{Introduction}
\label{sec:intro}
Following the recent \emph{Planck} data release, which has not detected any clear deviation from the predictions of the simplest single-field inflationary models, it is interesting to ask whether this should be taken as evidence that inflation was driven by a single field, or whether the data is still not good enough to discriminate between single and multiple-field models of inflation. In general, more complex models may have a better fit because they have more parameters which can be tuned, but they should be penalised if large amounts of parameter space are a bad fit to the data. 

We study the curvaton scenario, which is one of the most popular and widely studied multifield models of inflation. In the original scenario \cite{Lyth:2001nq, Enqvist:2001zp, Moroi:2001ct} (see also \cite{Mollerach:1989hu,Linde:1996gt}), the curvaton field generates the primordial curvature perturbation while the inflaton field drives inflation. By definition, the curvaton field is light and subdominant during inflation, but its energy density grows relative to the background energy density after inflation ends, due to the curvaton decaying much later than the inflaton. The curvaton scenario is perhaps the best known scenario for generating large non-Gaussianity (of the local shape), a potential signature which \emph{Planck} has constrained much more tightly than the \emph{WMAP} satellite but not detected.  Building on the original scenario, one can also consider mixed scenarios in which both the inflaton and curvaton significantly contribute to the power spectrum~\cite{Dimopoulos:2003az,Langlois:2004nn}, or the inflating curvaton scenario~\cite{Moroi:2005kz, Moroi:2005np, Ichikawa:2008iq, Dimopoulos:2011gb}, which changes the observational predictions of the model. 



We perform the first Bayesian analysis of the ``simplest'' curvaton scenario, consisting of an inflaton and a curvaton field both with quadratic potentials. This model was originally studied in 2002 \cite{Bartolo:2002vf}, an analysis of the observables produced was performed in 2008~\cite{Ichikawa:2008iq} and the predictions of the model were revisited following the BICEP2 data release in 2014 \cite{Byrnes:2014xua}. While those papers studied the predictions and parameter space of the model, there has been no work asking whether the data prefers the addition of the curvaton field. Interestingly, although this model is famous as a means of generating large non-Gaussianity, we find it is not the non-detection of non-Gaussianity which puts this model under observational pressure.

Recent papers have conducted comprehensive model selection analyses over parameter spaces from single-field inflationary potentials~\cite{Ade:2015oja, Ade:2013uln, Martin:2013nzq, Easther:2011yq, Giannantonio:2014rva} yet when reviewing the literature, there appears to be very little using the Bayesian paradigm over multi-field models such as the curvaton. We note that a Bayesian analysis with \emph{WMAP} data of a Peccei-Quinn field has previously been performed in 2004~\cite{Lazarides:2004we} which is effectively a physically motivated mixed inflaton-curvaton generation of the curvature perturbation. Best-fit parameter values over the $\{n_s,r\}$ space have also been previously obtained in~\cite{Ellis:2013iea} and sampling over the same space in~\cite{Moroi:2005np}, both for different choices of prior.

Conducting a model comparison on the curvaton presents a interesting test of how discerning the current data is, given that the choice of prior on parameters within the model can be difficult and misleading in certain cases. A particular choice of prior to carefully consider will be that of the vacuum expectation value of the curvaton $\sigma_*$, which we  choose to have either a Gaussian or flat prior --- a physically motivated or statistically relevant choice respectively --- and it is important to discuss how independent of these prior choices our conclusions are.      

In this work; we make the initial step in discerning how a model selection on the curvaton is to be accomplished; we perform a Bayesian analysis to ascertain how much the data may prefer, if at all, the single-field quadratic inflaton over adding a quadratic curvaton; and we quantify how useful future measurements on local-type primordial non-Gaussianity~\cite{Giannantonio:2011ya, Camera:2014bwa, Dore:2014cca, Leistedt:2014zqa} will be to favour or disfavour the curvaton.

\section{The curvaton scenario}

We shall firstly review the general predictions of the curvaton model with potential \cite{Bartolo:2002vf}
\begin{equation}
\label{potential}
\begin{split}
\ V(\phi,\sigma) &= \frac{1}{2}M^2\phi^2 + \frac{1}{2}m^2\sigma^2 \,,
\end{split}
\end{equation}
that can be used to make predictions of $4$ specific observational parameters: the power spectrum amplitude ${\cal P}_{\zeta}$, the scalar spectral index $n_s$, the tensor-to-scalar ratio $r$ and the non-linearity parameter $f_{\mathrm{NL}}$ which characterises the degree of local type non-Gaussianity. 

To find the total power spectrum, one simply sums the inflaton ${\cal P}^{\phi}_{\zeta}$ and curvaton ${\cal P}^{\sigma}_{\zeta}$ contributions 
\begin{equation}
\label{powerspec}
\begin{split}
\ {\cal P}_{\zeta} &= {\cal P}^{\phi}_{\zeta} + {\cal P}^{\sigma}_{\zeta} = \left.\frac{V^3}{12\pi^2 (\partial_{\phi}V)^2 M_P^4}\right|_* + \left.\frac{r_{\mathrm{dec}}^2V}{27\pi^2\sigma^2M_P^2}\right|_*\,,
\end{split}
\end{equation}
where `$*$' hereafter denotes the value as the current scale crossed the horizon during inflation, $M_P$ is the reduced Planck mass and $r_{\mathrm{dec}}$ is defined as~\cite{Lyth:2001nq}
\begin{equation}
\label{rdecdef}
\begin{split}
\ r_{\mathrm{dec}} = \left. \frac{3\rho_{\sigma}}{3\rho_{\sigma}+4\rho_{\phi}}\right|_{\mathrm{dec}}\,,
\end{split}
\end{equation}
where `$\mathrm{dec}$' denotes evaluation at curvaton decay time. In the original curvaton model one assumes that ${\cal P}^{\phi}_{\zeta}$ is negligible, we here are more general and allow an arbitrary contribution to the power spectrum from both fields, significantly increasing the available parameter space for curvaton-like scenarios~\cite{Dimopoulos:2003az,Langlois:2004nn}. 

The slow-roll parameters are
\begin{equation}
\label{slowroll}
\begin{split}
\epsilon \equiv \left. \frac{M_P^2}{2}\left(\frac{\partial_{\phi} V}{V}\right)^2 \right|_* & \simeq -\frac{\dot{H_*}}{H^2_*}\simeq  \frac{1}{2N_*}\,,
\qquad \eta_{\phi} \left.\equiv M_{P}^2 \frac{\partial^2_{\phi} V}{V} \right|_* \simeq \frac{1}{2N_*} \,, \\
 & \eta_{\sigma} \left. \equiv M_{P}^2 \frac{\partial^2_{\sigma} V}{V} \right|_* \simeq \frac{m^2}{M^2}\frac{1}{2N_*}\,,
\end{split}
\end{equation}
where $\partial^2_{a} \equiv \frac{\partial^2}{\partial a^2}$ and $N_*$ is the number of efolds during inflation from when the current scale crossed the horizon until the end of inflation (denoted by a subscript `${\rm end}$' below), which is given by
\begin{equation}
\label{nstarinf}
\begin{split}
\ N_*=\ln\left(\frac{a_{\mathrm{end}}}{a_*}\right) \simeq \frac{1}{M_P^2}\frac{\phi^2_*}{4} \,.
\end{split}
\end{equation}
Note that the curvaton provides a negligible contribution to $\epsilon$ because the inflaton dominates the total energy density as it drives inflation.

Using the slow-roll parameters one can find the spectral index~\cite{Wands:2002bn}
\begin{equation}
\label{specind}
\begin{split}
\ n_s -1 &= \frac{{\cal P}^{\sigma}_{\zeta}}{{\cal P}_{\zeta}}(-2\epsilon + 2\eta_{\sigma}) +\frac{{\cal P}^{\phi}_{\zeta}}{{\cal P}_{\zeta}}(-6\epsilon + 2\eta_{\phi})\,,
\end{split}
\end{equation}
derived by parameterising the degree to which the inflaton fluctuations contribute to the total (i.e.~observed) power spectrum amplitude $\mathcal{P}_{\zeta}$. For a given value of $\epsilon$, the tensor-to-scalar ratio can only be suppressed compared to its single-field value, and is given by \cite{Wands:2002bn}
\begin{equation}
\label{tensor}
\begin{split}
\ r = \frac{{\cal P}^{\phi}_{\zeta}}{{\cal P}_{\zeta}}16\epsilon\,.
\end{split}
\end{equation}

Local non-Gaussianity is a characteristic observable of the multifield inflation models which if detected would rule out all single-field inflationary models \cite{Creminelli:2004yq} (however see \cite{Martin:2012pe,Chen:2013aj, Mooij:2015yka}). 
The curvaton model can produce $10^5 \gtrsim f_{\mathrm{NL}} \geq -5/4$ partially depending on how late the curvaton decays, parametrised by $r_{\mathrm{dec}}$, where~\cite{Byrnes:2014xua}
\begin{equation}
\label{fnl}
\begin{split}
\ f_{\mathrm{NL}} = \frac{5}{12}\left(\frac{{\cal P}^{\sigma}_{\zeta}}{{\cal P}_{\zeta}}\right)^2\left(\frac{3}{r_{\mathrm{dec}}}-4-2r_{\mathrm{dec}}\right)\,,
\end{split}
\end{equation}
where $r_{\mathrm{dec}}=1$ acts as an attractor for the majority of points within its parameter space, because $r_{\mathrm{dec}}$ initially grows quickly and then asymptotes to unity once it dominates the background energy density, unless the curvaton decays early enough. We therefore find $f_{\mathrm{NL}}=-5/4$ in the limit of ${\cal P}^{\phi}_{\zeta}\ll {\cal P}_{\zeta}$ and a sufficiently late decaying curvaton. For the choice of priors we make (which are broad), we find the majority of non-Gaussianity values are in the region $10 \gtrsim f_{\mathrm{NL}} \geq -5/4$ and are hence not in tension with observations.

Implicit in all of the expressions above, $N_*$ has a general expression in all quadratic curvaton models. By assuming instantaneous inflaton decay~\cite{Byrnes:2014xua}  
\begin{equation}
\label{nstar}
\begin{split}
\ N_* = 58 + \frac{1}{4}\ln \frac{{\cal P}^{\phi}_{\zeta}}{{\cal P}_{\zeta}} -\frac{1}{4}N_{\mathrm{mat}}\,,  \end{split} 
\end{equation}
where $N_*$ is taken to correspond to the horizon crossing time of the \emph{Planck} pivot scale, $k_*=0.05 \mathrm{Mpc}^{-1}$. The expression includes $N_{\mathrm{mat}}$, the number of efolds with a matter-like equation of state caused by the curvaton field, while it oscillates after becoming dominant. If the curvaton never dominates the background energy density then $N_{\mathrm{mat}}=0$. 

The key value of $r_{\mathrm{dec}}=3/7$ corresponds to inflaton-curvaton equality, $\rho_{\phi}=\rho_{\sigma}$ at the curvaton decay time, at which time the Hubble parameter equals the curvaton decay rate,
$ \Gamma_{\sigma}\equiv H|_{\mathrm{dec}} $. We find two regimes are possible. Firstly, for $\Gamma_{\sigma}>H_{\rm eq.}$ the curvaton decays before it can dominate the total energy density and  $N_{\mathrm{mat}}=0$. Secondly, using the relation~\cite{Bartolo:2002vf}
\begin{equation}
\label{rdecgamma}
\begin{split}
\ r_{\mathrm{dec}} = \left(1+\left.\frac{4\rho_{\phi}}{3\rho_{\sigma}}\right|_{\mathrm{dec}}\right)^{-1} = \left(1+\sqrt{\frac{\Gamma_{\sigma}}{m}}\frac{8M^2_P}{\sigma^2_{*}}\right)^{-1}\,.
\end{split}
\end{equation}
we can obtain an expression for $N_{\mathrm{mat}}$ in the opposite regime $\Gamma_{\sigma}<H_{\rm eq.}$, which means that the curvaton decays after it dominates the energy density with $r_{\mathrm{dec}} \geq 3/7$ and
\begin{equation}
\label{nmat}
\begin{split}
\ N_{\mathrm{mat}} \simeq \frac{2}{3}\ln\frac{m\sigma^4_{*}}{36M^4_P\Gamma_{\sigma}}\,.
\end{split}
\end{equation}

We note that by setting ${\cal P}^{\phi}_{\zeta}={\cal P}_{\zeta}\gg {\cal P}^{\sigma}_{\zeta}$ in all of the expressions above one retrieves the predictions of the single-field quadratic inflaton which becomes a limiting case of the curvaton model in which no significant perturbations are generated by the curvaton but $N_*$ may still vary through equation (\ref{nstar}) and $N_{\mathrm{mat}}$, which is zero without the curvaton. The curvaton models are therefore found by setting arbitrary power spectra of the inflaton and curvaton in the equations above, yielding 3 interesting scenarios in particular:
\begin{enumerate}

\item{The \textbf{mixed inflaton-curvaton} scenario allows free choice of any sensible $M$ so that both inflaton and curvaton fluctuations may significantly contribute to the power spectrum. This introduces a free effective parameter ${\cal P}^{\phi}_{\zeta}/{\cal P}_{\zeta}$.}

\item{The \textbf{pure curvaton} scenario is a subset of the mixed inflaton-curvaton scenario, matching the original model proposed by~\cite{Lyth:2001nq}. The curvaton is the sole generator of the fluctuations, i.e.~${\cal P}^{\phi}_{\zeta}\ll{\cal P}_{\zeta}$.} 

\item{The \textbf{dominant curvaton decay} scenario is a subset of the pure curvaton scenario which fixes $r_{\mathrm{dec}}=1$ ($f_{\mathrm{NL}}=-5/4$) where the curvaton decays only after it has become dominant in the energy density. $r_{\mathrm{dec}}=1$ is the most likely value for both of the curvaton scenarios considered above, so this is an interesting case to consider.}

\end{enumerate}


It is important to note that in all of the scenarios above we assume that there are no surviving isocurvature modes, which in practise means that either the curvaton dominates before decay and/or that the Universe thermalises after the curvaton decay. Whether the isocurvature perturbations survive until the CMB forms depends on the microphysics of reheating and this is beyond the scope of our investigation. At any rate, the assumption that the Universe fully thermalises necessarily ensures full isocurvature decay. 

We note that throughout this paper we make the common simplifying assumption that the inflaton does not decay into the curvaton after inflation. However several papers have shown that even if only a small fraction of the inflaton field decays into the curvaton~\cite{Linde:1996gt, Linde:2005yw} then the amplitude of the curvaton perturbations $\delta \rho_{\sigma}/\rho_{\sigma}$ is suppressed and that the value of $f_{\mathrm{NL}}$ may be enhanced. Including this effect goes beyond the scope of this paper, but the increased freedom in $f_{\mathrm{NL}}$ may make the Planck non-Gaussianity constraints more effective in this case.

\begin{figure}[ht] 
 \centering
 \includegraphics[width=13cm]{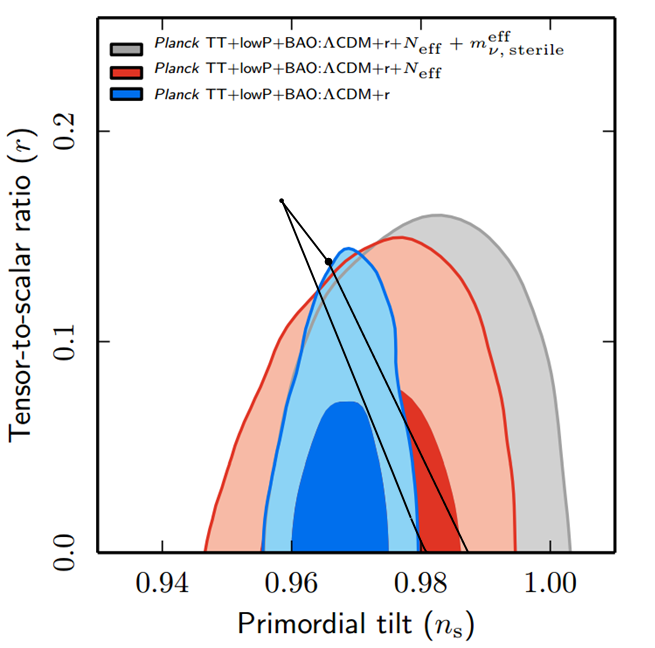}
\caption{\label{fig:1} The $\{n_s,r\}$ plot taken from~\cite{Ade:2015oja} outlining the regions where the larger dot is the inflaton single field inflaton limit at $N_*=58$, the black lines contain the mixed inflaton-curvaton and the $r=0$ values within this region belong to the pure curvaton scenarios. The left black line that corresponds to $N_* = 48$ at the top is essentially straight due to an approximate cancellation between the second and third terms of equation (\ref{nstar}). The right hand black line follows by maximising $N_*$ and $m$, whose upper limit we set as $m=M/2$ so that the curvaton does not roll significantly during inflation and in agreement with \cite{Byrnes:2014xua}. However the plot differs from the equivalent one in \cite{Byrnes:2014xua} because we here calculate the allowed range of $N_*$ as a function of all the model parameters, rather than assuming a fixed range of $N_*$. The \emph{Planck} data used in this paper corresponds to the blue contours and region. }
\end{figure}

Fig.~\ref{fig:1} illustrates the extrema boundaries available to the curvaton scenario over $\{n_s,r\}$ space in this paper. The single field inflaton is represented by the larger black point corresponding to $N_* = 58$ from equation (\ref{nstar}) which corresponds to a single point because we assume instantaneous decay of the inflaton. As shown in Fig.~\ref{fig:1}, the tensor-to-scalar ratio can be larger than the single-field value, in the case where the curvaton perturbations are negligible but its effect on the background evolution generates a large $N_{\mathrm{mat}}$, which can reduce $N_*$ to a minimum of $N_*=48$, making $\epsilon$ larger. The mixed inflaton-curvaton scenario is found from all points within the region defined between the three black lines. The predictions of the pure and dominant curvaton scenarios lie along the bottom of the plot ($r\simeq0$) between the two black lines.

We note that Fig.~\ref{fig:1} is derived using the single field consistency relation for $r$ -- this is used to evaluate $r$ with a pivot scale $k_*=0.002 {\rm Mpc}^{-1}$ -- but by contrast the scalar spectral index $n_s$ has no associated scale because the running is assumed to be $0$. This leaves us with a small fundamental error in the exact placement of the point over $\{n_s,r\}$ for the single field model because it predicts that both $n_s$ and $r$ change equally with the pivot scale ($n_s \propto r \propto 1/N_*$). The current constraint on the scalar running $dn_s/d\ln k=-0.003 \pm 0.007$~\cite{Ade:2015oja} is close to $0$, therefore we do not expect this effect to be significant to our overall analysis. In Fig.~\ref{fig:1} we calculate $n_s$ and $r$ at the usual Planck pivot scale of $0.05{\rm Mpc}^{-1}$ using (\ref{nstar}), which is the best choice for the spectral index but does introduce a small error into $r$ when it is large.

\section{Model Selection}
\subsection{Selection methods and prior ranges}
In the attempt to quantify the degree to which the current data is able to discern between models, one might expect the curvaton to be largely disfavoured against a single-field potential using the Bayesian approach. The curvaton boasts two additional parameters to its potential, plus the late time parameter $r_{\mathrm{dec}}$ with no further constraints, thereby losing `efficiency' in statistical fitting.  

We shall show that the curvaton performs much better than one might expect, highlighting some of the weakness in current data so far in discerning between models. In particular, the curvaton is not disfavoured against single-field quadratic inflation. To this end, we now briefly review the model selection methods used here.

We maximise the model likelihood $\mathcal{L}$ taking into account the data provided by \emph{Planck}~\cite{planckchains} to find $\Delta \chi^2_{eff}$
\begin{equation}
\label{chisq}
\begin{split}
\Delta \chi^2_{eff} &\equiv 2 [ \ln \mathcal{L}_{\max}(\mathcal{M}_1)-\ln \mathcal{L}_{\max}(\mathcal{M}_2) ] \,.
\end{split}
\end{equation}
Models which fare well in a comparison using $\Delta \chi^2_{eff}$ may not be as statistically favourable in the Bayesian paradigm due to a wastage of parameter space penalised by the evidence $\mathcal{E}$ with model parameters $k$
\begin{equation}
\label{evidence}
\begin{split}
\mathcal{E} &= \int d^kx \Pr (\boldsymbol{x}|\mathcal{M}) \mathcal{L}(data|\boldsymbol{x})\,,
\end{split}
\end{equation}
in ratio with another reference model and is typically quoted logarithmically. Once calculated, the $\ln (\mathcal{E} / \mathcal{E}_{ref})$ values must have some measure to correctly interpret their relevance in a model comparison, one such scale used in~\cite{Trotta:2008qt} is that of Jeffrey's which is shown in Table \ref{tab:0}. We are careful here to note that the Jeffrey's scale cannot be used as much more than a guideline, especially in instances where models appear to have similar evidences as it can be shown to be an incomplete comparison tool~\cite{Nesseris:2012cq}.
\begin{table}[tbp]
	\centering
	\begin{tabular}{|lr|c|}
		\hline
		$|\ln ( \mathcal{E} / \mathcal{E}_{ref} ) |$ & \textbf{Interpretation} \\
		\hline
		$< 1$  &   \emph{Inconclusive} \\ 
		 $1-2.5$  &  \emph{Weak evidence} \\
		 $2.5-5$  &  \emph{Moderate evidence} \\
		 $>5$  &  \emph{Strong evidence} \\
		\hline
	\end{tabular}
	\caption{\label{tab:0} The adapted Jeffrey's scale as used in~\cite{Trotta:2008qt} that assigns meaning to $\ln (\mathcal{E} / \mathcal{E}_{ref})$ values. }
\end{table}

An important concept with the introduction of Bayesian statistics is the choice of priors, which may be categorised as either physically motivated or non-informative (Jeffrey's prior)~\cite{jeffreyspriorcite}, the latter of which is proportional to the square root of the Fisher information determinant.

\begin{table}[tbp]
\centering
\begin{tabular}{|lr|c|}
\hline
\textbf{Model Scenarios} & \textbf{Prior ranges and constraints ($M_P=1$)} \\
\hline
Single field inflaton  &  $-4>\log_{10}(M)>-6.75$~\cite{Ade:2013uln} \\ 
\hline
Mixed inflaton-curvaton    & $-4>\log_{10}(M) > -15.3$~\cite{Ade:2013uln} \\
   &   $\log_{10} (M/2) > \log_{10}(m) > -39$~\cite{Bartolo:2002vf}  \\
   &   (flat or Gaussian) $|\sigma_*| < 0.01$~\cite{Langlois:2004nn} \\
   &   $\log_{10}(m) > \log_{10}(\Gamma_{\sigma}) > -39$~\cite{Bartolo:2002vf} \\
\hline
 Pure curvaton  &  Mixed inflaton-curvaton subset with extra constraint: \\ 
& ${\cal P}^{\phi}_{\zeta} \lesssim 0.01({\cal P}^{\phi}_{\zeta} +{\cal P}^{\sigma}_{\zeta}) = 0.01{\cal P}_{\zeta}$  \\ 
\hline
Dominant curvaton decay   &  Pure curvaton subset with extra constraint: \\
  &  $r_{\mathrm{dec}}=1$  \\
  &  $\Rightarrow \log_{10}(m) + 4\log_{10}(\sigma_*) - 4 > \log_{10}(\Gamma_{\sigma})$ (\ref{rdecgamma}) \\
\hline
$\Lambda$CDM concordance   &   $1.02>n_s>0.9$~\cite{Ade:2013uln}  \\
	&   $3.2>\ln (10^{10}{\cal P}_{\zeta}) >3.0$~\cite{Ade:2013uln} \\
	&   $r=0$~\cite{Ade:2013uln} \\
\hline
\end{tabular}
\caption{\label{tab:1} The list of prior ranges for all of the inflationary scenarios discussed in the text. 
In all models we set an overall prior of $10^{-7}>{\cal P}_{\zeta}>10^{-11}$, emulating~\cite{Ade:2013uln}. Unlike~\cite{Ade:2013uln} we do not set a prior on $N_*$, because it can be precisely calculated after all of the above model parameters have been specified. In the case of the curvaton scenarios we have placed an additional constraint of $10^{-7}>{\cal P}^{\sigma}_{\zeta}>10^{-11}$. }
\end{table}

We present our prior ranges for the scenarios discussed previously with our rationale in Table~\ref{tab:1}, where we shall now discuss our choices. The single field quadratic model in Table~\ref{tab:1} sets a non-informative logarithmic prior only on $M$, matching that of~\cite{Ade:2013uln} as we find that $N_*=58$ from equation (\ref{nstar}) for an instantaneous decay scenario (which means that the Hubble crossing value of $\phi$ is not a free parameter).

In all curvaton models we set a logarithmic prior over $M$ to match the single-field inflation model. In the mixed inflaton-curvaton scenario, $M$ is given a full range between its upper limit corresponding to the single field inflaton and its lower limit to the minimum $M$ allowable by the prior $10^{-7}>{\cal P}_{\zeta}>10^{-11}$ which we attribute globally to all models keeping only the region of parameter space that is compatible with basic structure formation requirements~\cite{Ade:2013uln}. In contrast, the pure curvaton must have a tighter upper limit on $M$ (by a factor $10$ compared to the mixed and single-field upper limits) resulting in a suppression of $r$. Because the choice of prior is logarithmic over the inflaton mass we find the effective parameter volume available to the pure curvaton scenario to be not much smaller than the mixed case.

In all relevant cases we also use a non-informative logarithmic prior over $m$ and obtain the upper limit from the requirement that the curvaton does not roll significantly until inflation ends, requiring $m<M/2$ which implies $\eta_\sigma<\eta_\phi/4$. In choosing the lower limit on $m$ and $\Gamma_{\sigma}$ one can allow for variations of baryogenesis to offer constraints however we choose the most conservative bound, which is big bang nucleosynthesis where $H_{nucl.} \simeq 10^{-39}M_P$.

We set a firm upper limit of $|\sigma_*|<10^{-2}M_P$ to stay within the regime where our equations for the curvaton are valid~\cite{Langlois:2004nn} and to avoid the inflating curvaton scenario. We have two choices of prior on $\sigma_*$. As the order of magnitude of $\sigma_*$ is much better known, the non-informative prior one can place is a flat prior. We also have a physically motivated alternative following~\cite{Enqvist:2012xn} to the obtain the equilibrium distribution of $\sigma_*$ in a de Sitter background
\begin{equation}
\label{probcurv}
\begin{split}
\Pr (\sigma_*) &\propto \exp \left(-\frac{8\pi^2}{3H^4_*}V(\sigma_{*})\right)\,.
\end{split}
\end{equation}
This distribution is obtained due to the relaxation of the vacuum expectation value at the equilibrium limit being normally distributed over $\sigma_* =0$ with a standard deviation $\propto M^2N_*/m$. We therefore carry out our analysis separately with both flat and Gaussian priors over $\sigma_*$ to assess the effect, if any, this may have on the evidences. Note that even when using a Gaussian prior on $\sigma_*$, we truncate the range of the Gaussian to enforce $|\sigma_*|<10^{-2}M_P$.

Finally, the prior we place over $\Gamma_{\sigma}$ in all scenarios is logarithmic where we find the ranges differ, setting an upper limit for the mixed and pure cases as $m$ so that the curvaton oscillates before it decays and an upper limit for the dominant curvaton scenario from fixing $r_{\mathrm{dec}}=1$ in equation (\ref{rdecgamma}). The lower limit of $\Gamma_{\sigma}$ in all $3$ curvaton scenarios follows from the constraint that the curvaton must decay before big bang nucleosynthesis.

\subsection{Calculating the evidences}~\label{sec:comp}
The evidence is far more computationally expensive to calculate than the maximum likelihood, particularly with large numbers of parameters, so other papers~\cite{Ade:2015oja, Ade:2013uln, Martin:2013nzq, Easther:2011yq} have implemented nested sampling techniques to compensate for the increase in parameter volume. By comparison direct Monte Carlo methods -- sampling across a grid of sufficiently narrow step size -- are still feasible as a way of estimating the evidence for relatively small parameter spaces such as $\{{\cal P}_{\zeta},n_s,r,f_{\mathrm{NL}}\}$ and we shall implement these methods in this paper.     

Initially, we choose the following computation ranges and bin widths:
\begin{equation*}
\begin{split}
\ 3.4>\ln(10^{10}{\cal P}_{\zeta})>2.8\,,
\qquad 1.02>n_s>0.9\,,
\qquad 0.5>r>0\,,
\qquad 30>f_{\mathrm{NL}}>-20\,,
\end{split}
\end{equation*}
\begin{equation*}
\begin{split}
\ \Delta \ln(10^{10}{\cal P}_{\zeta}) \simeq 0.006\,,
\qquad \Delta n_s \simeq 0.001\,,
\qquad \Delta r \simeq 0.005\,,
\end{split}
\end{equation*}
where by marginalising only over the $f_{\mathrm{NL}}$ measurement we reduce the effective number of parameters to be stored so that the data distribution can be represented in a multidimensional array and a Gaussian function is placed over $f_{\mathrm{NL}}=0.8$ with $\sigma(f_{\mathrm{NL}})=5.0$~\cite{Ade:2015ava}, therefore it does not have an associated bin width. The widths are roughly consistent with selecting histogram bin sizes following the Freedman-Diaconis rule~\cite{histogramref}, which takes into account the sample size of the \emph{Planck} Markov chains $\simeq 2 \times 10^{4}$ obtained from~\cite{planckchains}.


Implemented for all models in this paper, our numerical method was as follows: 

\begin{enumerate}
\item{Smooth the binned MCMC chains over $\{ \ln(10^{10}{\cal P}_{\zeta}), n_s, r \}$ by kernel.}
\item{Multiply the smoothed likelihood by a Gaussian $f_{\mathrm{NL}}$ function to obtain the likelihood over $\{ \ln(10^{10}{\cal P}_{\zeta}), n_s, r, f_{\mathrm{NL}} \}$.}	
\item{Obtain $10^6$ sampled points from our code to emulate a prior, where the single field was an effective delta function.}
\item{Sum the resultant likelihood from $2.$ over the prior mass to numerically approximate the evidence integral.}
\end{enumerate}

For a dramatic difference in computational efficiency, when obtaining the smoothed likelihood from the MCMC chains, we have chosen to ignore the covariance between the background cosmological parameters and our chosen parameters $\{ \ln(10^{10}{\cal P}_{\zeta}), n_s, r \}$ by foregoing marginalisation. This ignorance necessarily introduces an error into our calculation of the likelihood function over $\{ \ln(10^{10}{\cal P}_{\zeta}), n_s, r, f_{\mathrm{NL}} \}$ and therefore an error in the $\ln \mathcal{E}$ ratios. To test that this is small, we have compared our evidence ratios of monomial potentials, the simple Higgs inflationary potential and radiatively corrected Higgs potential to those obtained by \emph{Planck} 2013, \emph{Planck} 2015 and~\cite{Martin:2013nzq}.

The $\ln \mathcal{E}$ ratios for monomial potentials can be matched to \emph{Planck} 2013 in~\cite{Ade:2013uln} up to an error of $\mathcal{O}(0.1)$. \emph{Planck} 2015 use an alternative baseline model ($R^2$ inflation) to calculate their evidence ratios, however we are able to verify an error of $\mathcal{O}(0.1)$ when directly taking the ratios between our test monomial potentials. Finally, we find our error is also of $\mathcal{O}(0.1)$ when comparing our calculated evidence ratios to those obtained in~\cite{Martin:2013nzq} using monomial potentials, the simple Higgs inflationary potential and radiatively corrected Higgs potential.  


We present in Table~\ref{tab:3} $\Delta \chi^2_{eff}$ values and evidence ratios computed for each of the scenarios. The results are discussed in the next section. In table \ref{tab:4}  we calculate the required $1-\sigma$ error bar, $\sigma (f_{\mathrm{NL}})$, which observations are required to reach for each of the curvaton scenarios to have $\ln (\mathcal{E}/\mathcal{E}_{ref}) \simeq 0$ against $\Lambda$CDM. We do this for two fiducial values of $f_{\mathrm{NL}}\neq0$, firstly for the attractor value of ($r_{\mathrm{dec}}=1$) $f_{\mathrm{NL}}=-5/4$ and secondly for the current $2-\sigma$ upper bound value of $f_{\mathrm{NL}}=10.8$~\cite{Ade:2015ava}. These results assume that apart from a detection of $f_{\mathrm{NL}}$, all other cosmological data including the observational constraints on the spectral index and $r$ would remain the same.
\begin{table}[tbp]
	\centering
	\begin{tabular}{|lr|c|}
		\hline
		\textbf{Model Scenarios}   &  $\Delta \chi^2_{eff}$ & $\ln (\mathcal{E}/\mathcal{E}_{ref})$ \\
		\hline
		Single field inflaton    & $6.6$ & $-5.2$  \\
		Mixed inflaton-curvaton  ($\sigma_*$ flat prior)     &  $3.7$ & $-4.9$  \\
		Mixed inflaton-curvaton  ($\sigma_*$ Gaussian prior (\ref{probcurv}))     &  $3.7$ & $-4.6$  \\
		Pure curvaton  ($\sigma_*$ flat prior)         & $4.7$  & $-4.7$    \\
		Pure curvaton  ($\sigma_*$ Gaussian prior)         & $4.7$  & $-4.6$    \\
		Dominant curvaton decay ($\sigma_*$ flat prior)    & $4.7$ & $-4.6$   \\
		Dominant curvaton decay ($\sigma_*$ Gaussian prior)    & $4.7$ & $-4.4$   \\
		\hline
	\end{tabular}
	\caption{\label{tab:3}The $\ln (\mathcal{E}/\mathcal{E}_{ref})$ and $\Delta \chi^2_{eff}$ comparison statistics in the form of ratios against the $\Lambda$CDM refererence model over $\{{\cal P}_{\zeta},n_s,r,f_{\mathrm{NL}}\}$ space with the \emph{Planck} TT+lowP+BAO dataset as in~\cite{Ade:2015oja}.}
\end{table}
\begin{table}[tbp]
	\centering
	\begin{tabular}{|lr|c|}
		\hline
		\textbf{Model Scenarios} & fiducial  $f_{\mathrm{NL}}$  & $\sigma (f_{\mathrm{NL}})$ \\
		\hline
		Mixed inflaton-curvaton ($\sigma_*$ flat prior)  &   $-5/4$    & $0.4$ \\
		Mixed inflaton-curvaton ($\sigma_*$ Gaussian prior)  &   $-5/4$    & $0.4$  \\
		Pure curvaton ($\sigma_*$ flat prior)  &   $-5/4$   & $0.4$ \\
		Pure curvaton ($\sigma_*$ Gaussian prior)  &   $-5/4$   & $0.4$ \\
		Dominant curvaton decay ($\sigma_*$ flat prior)  &  $-5/4$   & $0.4$ \\
		Dominant curvaton decay ($\sigma_*$ Gaussian prior)  &  $-5/4$   & $0.4$ \\
		Mixed inflaton-curvaton ($\sigma_*$ flat prior)  &   $10.8$    & $2.8$ \\
		Mixed inflaton-curvaton ($\sigma_*$ Gaussian prior)  &   $10.8$    & $2.6$ \\
		Pure curvaton ($\sigma_*$ flat prior)  &   $10.8$   & $2.4$ \\
		Pure curvaton ($\sigma_*$ Gaussian prior)  &   $10.8$   & $2.6$  \\
		\hline
	\end{tabular}
	\caption{\label{tab:4} A list of required $f_{\mathrm{NL}}$ measurements with $\sigma({f_{\mathrm{NL}}})$ precisions which observations are required to reach for each of the curvaton scenarios to have $\ln (\mathcal{E}/\mathcal{E}_{ref}) \simeq 0$ against $\Lambda$CDM. The measurement at $f_{\mathrm{NL}}=10.8$ reflects the current $2-\sigma$ upper limit from observation~\cite{Ade:2015ava} and $f_{\mathrm{NL}}=-5/4$ would be a strong indication of the curvaton as it is derived from the case where $r_{\mathrm{dec}}=1$ and the inflation perturbations are subdominant compared to the curvaton perturbations. }
\end{table}

One can find by (arbitrarily) setting $f_{\mathrm{NL}}=0$ for all models and re-sampling that the $\ln (\mathcal{E}/\mathcal{E}_{ref})$ values are unchanged up to $\mathcal{O}(0.1)$, which means that despite the curvaton scenario being a famous method of generating a potentially large non-Gaussianity, the observational constraint on $f_{\mathrm{NL}}$ does not significantly disfavour the model or change the Bayesian evidence. However, in sharp contrast, by changing instead the power spectrum for all models, fixing them to be the same as $\Lambda$CDM concordance with $3.2>\ln (10^{10}{\cal P}_{\zeta})>3.0$, we find an increase in the $\ln (\mathcal{E}/\mathcal{E}_{ref})$ values by approximately 4.

With the especially tight error bar on ${\cal P}_{\zeta}$ we expect to find at least one finely tuned parameter in all of the models we study. Such parameters may have wide prior ranges but only specific values within those ranges will correspond to a favoured region by the data, e.g.~$\log_{10}(M/M_P) \simeq -5.2$ for the single field fits the power spectrum well, but the observational error bar is less than $2\%$ of the full range that is allowable by our globally imposed $10^{-7}>{\cal P}_{\zeta}>10^{-11}$ prior.

Altering the large ranges of both $m$ and $\Gamma_{\sigma}$ by shortening the lower bound from nucleosynthesis ($10^{-39}M_P$) to $10^{-30}M_P$ had minimal effect on the mixed curvaton $\log$ evidence -- shifting its magnitude by just $0.1$ -- therefore we find with our results against this choice of prior is quite robust.

We set $M$ to its favoured value for the curvaton model in section~\ref{sec:comp} and found essentially no change in the evidence, whereas setting the same value for the single field suppressed most of the disfavoured region of its parameter volume and therefore the $\log$ evidence shifted from $-5.2$ to $-1.3$. One can explain this difference between models because by imposing a logarithmic prior on $M$, we find that most of the parameter volume of the mixed curvaton has ${\cal P}^{\sigma}_{\zeta}>0.99{\cal P}_{\zeta}$ i.e.~the curvaton contributes most to the admixture of the perturbations, therefore the parameters that determine its magnitude, $r_{\mathrm{dec}}$ and $\sigma_*$ are those most affected by fine tuning. 

We further note that because the majority of points in all of the curvaton scenarios have $r_{\mathrm{dec}}=1$, the finely tuned parameter of the curvaton model is essentially its vacuum expectation value $\sigma_*$. To prove that $\sigma_*$ was fine tuned, we fixed both $\log_{10}(M/M_P) \simeq -6.1$ and $\sigma_* = 0.01$, corresponding to the region in the parameter volume of the mixed curvaton that is favoured in the data by ${\cal P}^{\sigma}_{\zeta}$ and we found a shift in the $\log$ evidence from $-4.9$ to $-1.9$ as expected. The compensation for the extra parameters in the evidence can therefore be traced back to the choice of range in $\sigma_*$, but is essentially independent of its prior over the given range -- Gaussian or flat -- because the $\log$ evidence ratios do not vary significantly with this choice.

Performing the same analysis with the chains from the \emph{Planck} 2013 release~\cite{Ade:2013uln} we found the evidence ratio of the mixed inflaton-curvaton scenario to be more disfavoured at $-5.2$ whereas the single field model was marginally preferred at $-5.0$. Following on from this we also obtained $\Delta \chi_{eff}^2$ and $\ln (\mathcal{E}/\mathcal{E}_{ref})$ values of the single field and curvaton scenarios using the BICEP/Keck/\emph{Planck} data from~\cite{Ade:2015oja} and one can see that the curvaton improves against the single-field inflaton, where differences in the $\log$ evidence between the two models are generally between $1.6 - 2.3$.    
\begin{table}[tbp]
	\centering
	\begin{tabular}{|lr|c|}
		\hline
		\textbf{Model Scenarios}   &  $\Delta \chi^2_{eff}$ & $\ln (\mathcal{E}/\mathcal{E}_{ref})$ \\
		\hline
		Single field inflaton    & $10.4$ & $-7.0$  \\
		Mixed inflaton-curvaton  ($\sigma_*$ flat prior)     &  $2.7$ & $-5.4$  \\
		Pure curvaton  ($\sigma_*$ flat prior)         & $4.8$  & $-4.7$    \\
		Dominant curvaton decay ($\sigma_*$ flat prior)    & $4.8$ & $-4.7$   \\
		\hline
	\end{tabular}
	\caption{\label{tab:5}The $\ln (\mathcal{E}/\mathcal{E}_{ref})$ and $\Delta \chi^2_{eff}$ comparison statistics in the form of ratios against the $\Lambda$CDM refererence model over $\{P_{\zeta},n_s,r,f_{\mathrm{NL}}\}$ space with the \emph{Planck} TT+lowP+BKP dataset as in~\cite{Ade:2015oja}.}
\end{table}
\subsection{Discussion}~\label{sec:discussion}
It is clear from looking at the evidence ratios in Table~\ref{tab:3} that the data disfavours both the curvaton and quadratic single-field inflation model significantly with respect to the reference $\Lambda$CDM concordance model where all models listed have close to ``\emph{strong evidence}'' against in accordance with Table~\ref{tab:0}, where the best fitting model according to its evidence appears to be the dominant curvaton by a fraction that is not significant according to the Jeffrey's scale. 

The data appears to yield no discernible difference in the evidence between the single field inflaton and any of the curvaton models, the $\Delta\chi^2_{eff}$ values indicate that the curvaton scenarios fit the data better but are penalised due to their parameter space wastage. The mixed inflaton-curvaton scenario has the better fit compared to the other two scenarios from its $\Delta\chi^2_{eff}$ value due to the fact that it may cross deeper into the favoured region over $\{n_s,r\}$ space with a larger tensor-to-scalar ratio, as can be seen from Fig.~\ref{fig:1}.

Performing our analysis with the BICEP/Keck data in section~\ref{sec:comp} finds a relative improvement in the evidence of the curvaton, where on the Jeffrey's scale, all scenarios are categorised to prefer single-field inflation with ``\emph{Weak evidence}''. However, evidence for all curvaton scenarios relative to $\Lambda$CDM is almost the same as using \emph{Planck} data alone, the only significant change is the tighter constraint on $r$ making the single-field model more disfavoured. 

At the end of section~\ref{sec:comp} we calculated the changes that fixing the power spectrum predictions to those of $\Lambda$CDM have on the $\log$ evidence ratios and it is found in Table \ref{tab:4} that in all cases the values shift by approximately $4$. This is an interesting result as it not only demonstrates that the parameter with the largest influence on model selection for both the curvaton and single field inflaton is ${\cal P}_{\zeta}$, but also because the shift is roughly uniform for all models, it implies that the difference in the evidences between curvaton and single-field inflaton predominately derives from their differing parameter spaces over $n_s$ and $r$ space (but not $f_{\rm NL}$ which is too weakly constrained to have much effect). 
 
The choice of $\sigma_*$ prior appears to have little effect on the evidences of Table~\ref{tab:3}, the pure and mixed curvaton scenarios still are disfavoured versus $\Lambda$CDM concordance with the same value and it is only the dominant curvaton scenario that appears to become slightly more favoured with the Gaussian prior. The logarithmic priors over a wide range of $M$ and $m$ largely reduce the effect that the constraint ${\cal P}^{\phi}_{\zeta} \lesssim 0.01{\cal P}_{\zeta}$ has in the pure curvaton scenario, making it unsurprising that there is little difference between the evidence ratios of the pure and mixed cases. We also found almost no sensitivity to changes in the minimum curvaton decay rate, increasing the lower bound on $\Gamma_{\sigma}$ by ten orders of magnitude led to almost no change in the Bayesian evidence ratios.

In Table~\ref{tab:4} we find the required measurements and their associated precisions $\sigma (f_{\mathrm{NL}})$ to make the curvaton scenario have the same evidence as the $\Lambda$CDM concordance model, 
assuming that all other  observational data remains the same. We additionally assume that the measured value of $f_{\mathrm{NL}}$ is centred on the fiducial value, with an error bar given in the right column of Table~\ref{tab:4}. If $f_{\mathrm{NL}}$ is as large as currently allowed at $2-\sigma$, then Euclid should reach sufficient accuracy to favour the curvaton scenario, but in the limit of the dominant curvaton scenario which predicts $f_{\mathrm{NL}}=-5/4$ an error bar significantly below $1$ will be required, which is forecast to be achievable in the longer term \cite{Giannantonio:2011ya,Camera:2014bwa,Dore:2014cca,Leistedt:2014zqa}.
Although a much higher precision on the measurement of $f_{\mathrm{NL}}$ is required if $f_{\mathrm{NL}}=-5/4$, the required significance of the detection is about 3-$\sigma$, which is lower than the detection significance of $3.5$ to $4.5-\;\sigma$ if $f_{\mathrm{NL}}=10.8$. The reason is because a large part of the curvaton prior space corresponds to the dominant curvaton scenario, with $r_{\mathrm{dec}}=1$ and $f_{\mathrm{NL}}=-5/4$. In contrast, only a small region of parameter space corresponds to $f_{\mathrm{NL}}\simeq10$. This also means that if future data does detect $f_{\mathrm{NL}}=-5/4$ we would have a strong motivation to believe this comes from the dominant curvaton scenario, but if $f_{\mathrm{NL}}\simeq10$ was detected one would want to also explore other scenarios capable of generating non-Gaussianity, to see if any model predicted this value in a large range of its parameter space.

Near the end of section~\ref{sec:comp} we also made a comparison between the impact of the recent \emph{Planck} 2015 release to the previous data from 2013 on the evidence. We find that the mixed curvaton becomes marginally more favoured against the single field inflaton by using the 2015, shifting from $\ln(\mathcal{E}/\mathcal{E}_{ref}) = -5.2$ to $-4.8$. This slight change we deduce is from the tighter constraints placed over the tensor-to-scalar ratio in 2015, mildly favouring lower $r$ regions than in the 2013 release. 

Notice that by applying the prior $|\sigma_*|<10^{-2}$ (a small range compared to the inflaton's initial field value) we do not attempt to quantify whether the curvaton scenario is likely within a multi-field context. We are instead asking whether the curvaton scenario is preferred within the context of a Lagrangian with two quadratic potentials where one has been given a large initial field value to drive inflation, while the second has a small field value and small decay rate to potentially act as a curvaton. Within this context, we find that \emph{Planck} data is inconclusive but the addition of BICEP/Keck data prefers the curvaton scenario.   

\section{Conclusions}~\label{sec:conclusions}
We find that both single-field quadratic inflation and the curvaton scenario (added to an inflaton field with quadratic potential) are strongly disfavoured relative to a $\Lambda$CDM reference model, which has Gaussian perturbations and no tensor perturbations. However the data is not good enough to discriminate between single-field quadratic inflation and the curvaton scenario, despite the curvaton scenario having three additional parameters. The priors are listed in Table \ref{tab:1}.

We studied three variants of the curvaton scenario, $1)$ the original ``pure'' scenario in which the inflation perturbations are neglected, $2)$ a subset of this case where the curvaton dominates the background energy density before it decays and $3)$ the mixed inflaton-curvaton scenario in which the perturbations of the inflaton are allowed to be large. The Bayesian evidence was similar for all three cases, probably because the majority of the prior space matches the dominant curvaton scenario. The mixed scenario has the best fit, as can be seen by eye from Fig.~\ref{fig:1}, corresponding to a value of $r\simeq0.06$. However only a small part of the available parameter space (were the inflation mass is close to its required value in the single field case) generates a non-negligible value of $r$. In most of the allowed parameter space, $r$ is negligibly small for the mixed scenario, as it is by definition in the pure and dominant curvaton scenarios. 

The Bayesian evidence ratios are not very sensitive to our choice of priors. We consider both a uniform and a Gaussian prior over the initial curvaton value, $\sigma_*$, and find very little change in the results. In both cases we require $|\sigma_*|<0.01 M_P$. Interestingly, we have found that the possibility of large non-Gaussianity in the curvaton scenario does not make it disfavoured, since if one drops the observational constraint on $f_{\mathrm{NL}}$ the evidence ratios hardly change. We also apply the priors $10^{-7}>{\cal P}_{\zeta}>10^{-11}$ and $10^{-7}>{\cal P}^{\sigma}_{\zeta}>10^{-11}$. For simplicity we assume that the inflation decays instantaneously after inflation ends, dropping this requirement would leave the inflation scenario with two free parameters and the curvaton with five parameters.  

The degeneracy in the evidences of single field and the curvaton in this paper demonstrate a situation where the Jeffrey's scale has essentially failed to find a preferred model given the data, despite some penalties included for wastage of parameter space. An important next step in order to break the degeneracy might be to find the Bayesian complexity of both models as in~\cite{Martin:2013nzq}. We expect such an analysis would be more punishing to the curvaton due to the additional parameters it requires. 

We find the required measurements and their associated precisions $\sigma (f_{\mathrm{NL}})$ to render the curvaton scenarios the same as the $\Lambda$CDM concordance model in the evidence, assuming that all other observational data remains the same. The conclusions from this analysis are twofold: firstly, if $f_{\mathrm{NL}}$ is as large as currently allowed at $2-\sigma$ from \emph{Planck} 2015, then Euclid should reach sufficient accuracy to favour the curvaton scenario and secondly, in the limit of the dominant curvaton scenario which predicts $f_{\mathrm{NL}}=-5/4$ an error bar significantly below $1$ will be required, which is forecast to be achievable in the longer term \cite{Giannantonio:2011ya,Camera:2014bwa,Dore:2014cca,Leistedt:2014zqa}. Unless $f_{\rm NL}$ is significantly larger than unity, the error bar on $f_{\rm NL}$ needs to be decreased by over an order of magnitude before non-Gaussianity has a significant effect on model selection. 

We have shown it is possible to find a simple multi-field model that is indistinguishable from the single-field quadratic inflaton using the Bayesian evidence combined with the \emph{Planck} 2015 data, despite having thee additional model parameters. The curvaton scenario is favoured when using the BICEP/Keck/\emph{Planck} data. We suggest that this result may be viewed in two ways: firstly, we may further strengthen the incentive to improve the current constraint on primordial non-Gaussianity as a tool to distinguish between inflationary potentials and secondly, one may take this result as motivation to explore the parameter space and Bayesian evidence of other multi-field models in future work.    

\acknowledgments

RH was supported by a Royal Society Summer Research Studentship for the main duration of this work. CB is supported by a Royal Society University Research Fellowship. The authors wish to thank Tommaso Giannantonio, Antony Lewis, Andrew Liddle, Jerome Martin, Hiranya Peris, Christophe Ringeval and especially Vincent Vennin for helpful discussions.

\bibliographystyle{JHEP}
\bibliography{my_biblio}

\end{document}